# Observation of Gigahertz Topological Valley Hall Effect in Nanoelectromechanical Phononic Crystals

Qicheng Zhang[1, †], Daehun Lee[2, †], Lu Zheng[2], Xuejian Ma[2], Shawn I. Meyer[2], Li He[1], Han Ye[3], Ze Gong[4], Bo Zhen[1], Keji Lai[2,*], A. T. Charlie Johnson[1,4*]

[1] Department of Physics and Astronomy, University of Pennsylvania, Philadelphia 19104, USA

[2] Department of Physics, University of Texas at Austin, Austin, TX, USA

[3] State Key Laboratory of Information Photonics and Optical Communications, Beijing University of Posts and Telecommunications, Beijing, China

[4] Department of Materials Science and Engineering, University of Pennsylvania, Philadelphia, PA 19104, USA

[†] These authors contributed equally to this work
* E-mails: kejilai@physics.utexas.edu; cjohnson@physics.upenn.edu

## Abstract

Topological phononics offers numerous opportunities in manipulating elastic waves that can propagate in solids without being backscattered. Due to the lack of nanoscale imaging tools that aid the system design, however, acoustic topological metamaterials have been mostly demonstrated in macroscale systems operating at low (kilohertz to megahertz) frequencies. Here, we report the realization of gigahertz topological valley Hall effect in nanoelectromechanical AlN membranes. Propagation of elastic wave through phononic crystals is directly visualized by microwave microscopy with unprecedented sensitivity and spatial resolution. The valley Hall edge states, protected by band topology, are vividly seen in both real- and momentum-space. The robust valley-polarized transport is evident from the wave transmission across local disorder and around sharp corners, as well as the power distribution into multiple edge channels. Our work paves the way to exploit topological physics in integrated acousto-electronic systems for classical and quantum information processing in the microwave regime.

The hallmark of topological phases of matter is the existence of nontrivial boundary states, which are protected by the global property of bulk energy bands and thus robust against local perturbations[1-3]. While originally developed for electron waves in condensed matter systems, topological band theory soon expanded into other wave systems, leading to the rapid advance of topological mechanics[4,5], photonics[6,7], and phononics[8,9] in the past decade. In particular, transport through edge channels in topological phononic systems is protected from backscattering due to structural disorders, fabrication imperfections, and environmental changes[10-20]. As a result, the pursuit of artificial elastic structures with novel band topology is not only intellectually attractive but also technologically relevant, especially in configurations that are suitable for integrated circuit applications.

The realization of a topological phase of matter requires coordinated efforts in theoretical calculation, experimental implementation, and systematic characterization. In electronic materials, this is achieved through first-principles calculations, material synthesis, macroscopic transport, momentum-space ($k$-space) mapping, and microscopic imaging[2]. The same design principle also applies to acoustic metamaterials[8,9,21], whose elastic moduli or mass densities vary periodically on a scale comparable to the acoustic wavelength. To date, studies of topological phononic crystals PnCs have been limited to relatively low frequencies. In the kHz range, the sonic design can be implemented by periodic arrangement of centimeter-scale meta-molecules (rods, rings, prisms…), and the local pressure field of the sound wave is measured by a microphone[13-15,18]. For structures operating in the MHz regime, the local displacement field of the elastic wave is usually detected by a scanning laser interferometer[16,17,19,20]. In contrast, topological PnCs are much less explored in the Ultra High Frequency (UHF, 0.3 – 3 GHz) and Super High Frequency (SHF, 3 – 30 GHz) bands that are of crucial importance for wireless communication[22], sensing technology[23], and universal quantum interconnects[24]. One major hurdle lies in the difficulty to probe phononic structures fabricated on monolithic substrates. Given the micrometer-scale GHz acoustic wavelength in solids, optical vibrometers[16,17,19,20] with diffraction-limited spatial resolution (~ 1 µm) are not generally applicable in the UHF/SHF regime. The development of a nanoscale acoustoelectric probing technology is therefore highly desirable for the advance of GHz topological phononics.

In this work, we report the realization of gigahertz topological valley Hall insulators and the observation of robust valley-polarized edge states in suspended AlN phononic crystals. Using a newly developed technique of transmission-mode microwave impedance microscopy (TMIM)[25,26], we visualize the wave pattern across gapped and gapless PnCs on microfabricated freestanding membranes. Edge transport channels are vividly seen at the interface between two domains with opposite valley Chern numbers and the band dispersion is consistent with reciprocal-space analysis using fast Fourier transform (FFT). We show that the edge transport is robust against local disorder at the domain wall and sharp corners in Z-shaped channels, highlighting its topological protection against backscattering events. Topology-protected wave steering is also achieved in a beam-splitter design. Our results represent the first successful demonstration of topologically nontrivial phases in integrated phononic circuits with an operation frequency in the microwave regime, which is promising for both classical information technology and quantum computation systems.

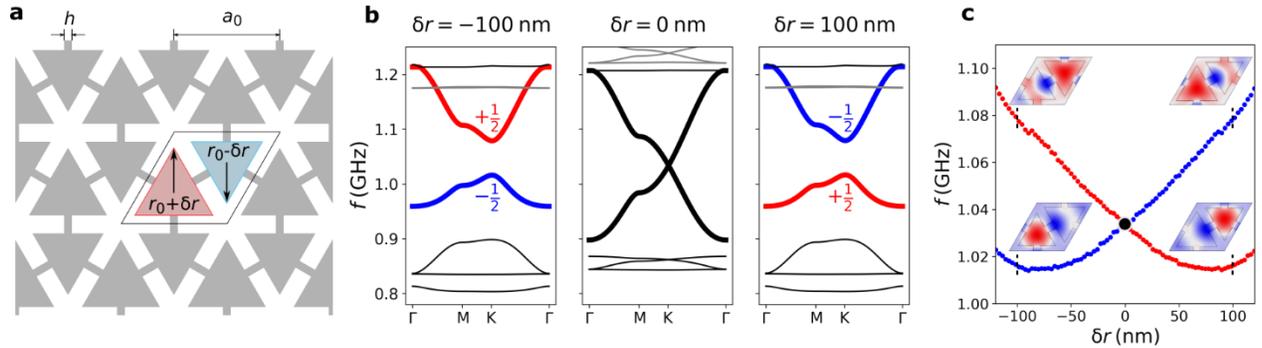

**Fig. 1 | Phononic crystal design and calculated band structures. a**, Schematic of the metamaterial design. The white regions are etched away from a freestanding AlN membrane. The rhombus indicates the unit cell, which consists of two equilateral triangles and connecting bridges. Parameters in the figure are $a_0 = 4.9$ μm, $r_0 = 2$ μm, $h = 0.36$ μm, and $\delta r = +100$ nm. Only $\delta r$ is varied among different samples. **b**, Left to right: Simulated band structures with $\delta r = -100$ nm, 0 nm, and 100 nm. The topological bands relevant to this work are highlighted. Dirac dispersion is observed in the middle structure with $\delta r = 0$. The valley Chern numbers near the band extrema are labeled for the gapped structures. **c**, Eigenfrequencies at the top/bottom band edges of interest as a function of $\delta r$. The black dot denotes the band closing point and valley Hall phase transition at $\delta r = 0$. The insets depict the eigenmodes of piezoelectric potential within a unit cell at the band extrema for $\delta r = \pm 100$ nm.

The topological nanoelectromechanical structure implemented in this work is based on the phononic analogue of the valley Hall effect (VHE)[27]. Valley pseudospin[28] labels the degenerate energy extrema in momentum space, whose topological nature stems from the nontrivial Berry curvature localized at these extremal points of the acoustic band structure. As depicted in Fig. 1a, snowflake-like[29,30] regions are etched away from a freestanding AlN membrane[31,32]. The rhombic primitive cell consists of two sublattice sites (equilateral triangles, radii of circumcircles $r_0 \pm \delta r$) and linking bridges. Fig. 1b shows the calculated band structures using finite-element modeling (see Methods). The dependence of eigenfrequencies as a function of various dimensions of this snowflake pattern is shown in Supplementary Information S1. When $\delta r = 0$, the structure can be viewed as an acoustic analogue of graphene, with $C_{6v}$ symmetry (sixfold rotations about the snowflake center and mirror symmetry about the vertical planes containing a basis vector of the honeycomb lattice), resulting in time-reversal-symmetry-protected Dirac dispersion near the $K$ and $K'$ points of the Brillouin zone. For $\delta r \neq 0$, the symmetry of the PnC reduces to $C_{3v}$, and the broken inversion symmetry lifts the degeneracy at the Dirac point[14,15,18,20,33-35]. Fig. 1c shows the $\delta r$-dependent bandgaps, as well as the piezoelectricity induced electrical potential profiles at the band edges for $\delta r = \pm 100$ nm. The topological phase transition across $\delta r = 0$ can be captured by an effective **k·p** Hamiltonian[14]:

$$H_{K(K')} = v_D\big(\kappa_x \sigma_x + \kappa_y \sigma_y + m v_D \sigma_z\big), \tag{1}$$

where $v_D$ is the Dirac velocity, $\kappa_{x,y,z}$ the momentum deviation from the $K$ ($K'$) point, $\sigma_{x,y,z}$ the Pauli matrices, and $m$ the mass term introduced by the broken inversion symmetry. This massive Dirac Hamiltonian produces nontrivial Berry curvatures at the two valleys, each can be integrated into a topological charge, i.e., a valley Chern number $C_{K(K')} = \frac{1}{2}\,\mathrm{sgn}(m)$.[14] Here sgn($m$) refers to the sign of $m$. It should be noted that because of the large momentum separation between the inequivalent $K$ and $K'$ valleys, intervalley scattering is often significantly suppressed. As is seen below, the quantization of $C_{K(K')}$, arising from mass inversion and the assumption of inhibited intervalley scattering, plays crucial roles in the emergence of gapless VHE edge states.

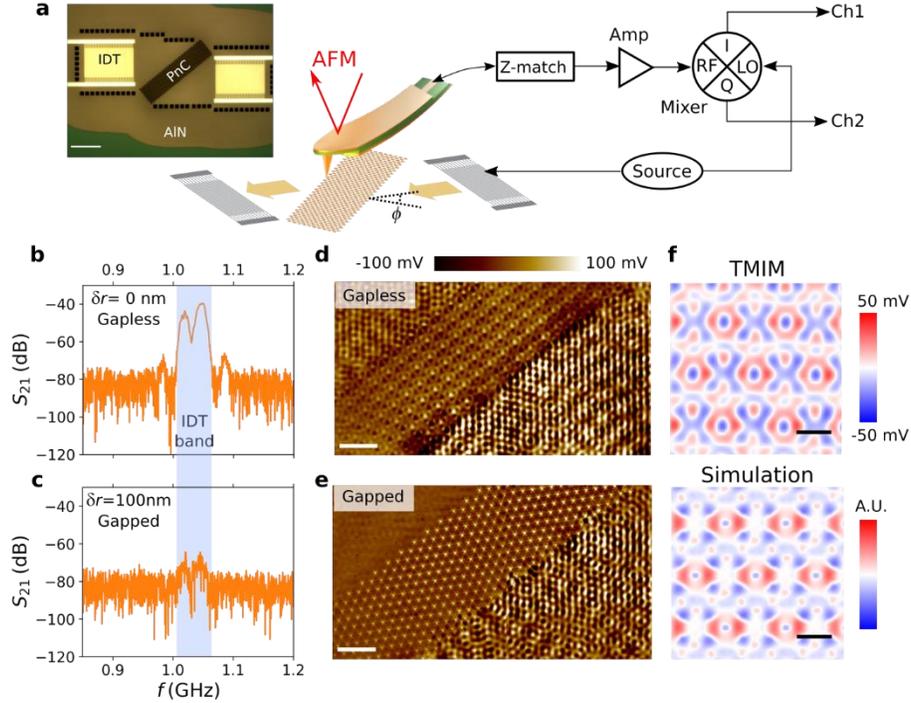

**Fig. 2 | Characterization of gapless and gapped phononic crystals. a**, Left: Picture of the phononic device fabricated on a suspended AlN membrane. Right: Schematic of the TMIM setup and the acoustic device with input/output IDTs shown in gray. **b**, Transmission coefficient as a function of excitation frequency for the gapless device with $\delta r = 0$ nm and **c**, gapped device with $\delta r = 100$ nm. The IDT passband is shaded in blue in both plots. The corresponding TMIM images at $f = 1.045$ GHz are shown in **d** and **e**, respectively. **f**, TMIM image inside the gapless PnC in (**d**), plotted in a different false-color map. The lower image shows simulated piezoelectric potential profile for the same field of view. Scale bars are 40 μm in (**a**), 20 μm in (**d**, **e**) and 5 μm in (**f**).

The design of our PnC is geometric in nature and fully compatible with standard nanofabrication processes[31,32]. The inset of Fig. 2a shows the nanoelectromechanical device fabricated on a freestanding 800-nm-thick *c*-axis polycrystalline AlN film. Future experiments could incorporate Sc-doped AlN thin films[36] that will further improve the electromechanical coupling. We emphasize that, in comparison with a prior work[16], the relatively thick and stiff film and the small dimension of the triangular artificial atoms are crucial for the realization of GHz characteristic frequencies. The PnC is patterned by electron-beam lithography and plasma etching. The interdigital transducers (IDTs) used to excite acoustic waves in the film are formed by depositing 45-nm-thick Al on the piezoelectric AlN. Visualization of the GHz acoustic waves is performed by TMIM on an atomic-force microscope (AFM) platform[25,26]. All measurements are performed under ambient temperature and pressure. As illustrated in Fig. 2a, the acoustic wave is launched by the emitter IDT. The shielded cantilever probe[37] behaves as a microwave receiver to pick up the ~ 1 GHz piezoelectric potential. Through the 50-$\Omega$ impedance-match section, the signal is amplified and demodulated by an in-phase/quadrature (I/Q) mixer, using the same microwave source that drives the emitter IDT. The signals at the RF and LO ports of the mixer can be expressed as $V_{RF} \propto e^{i(\omega t - kx)}$ and $V_{LO} \propto e^{i(\omega t + \theta)}$, where $\omega$ is the angular frequency, $k$ the acoustic wave vector, and $\theta$ the mixer phase, respectively. The two output channels are therefore $V_{Ch1} \propto \text{Re}(V_{RF}V_{LO}^*) = \cos(kx + \theta)$ and $V_{Ch2} \propto \text{Im}(V_{RF}V_{LO}^*) = -\sin(kx + \theta)$. The complex-valued TMIM signal $V_{Ch1} + i * V_{Ch2}$ provides a phase-sensitive measurement on the local displacement field of the elastic wave. For simplicity, we will only present one channel below unless otherwise specified. The unique properties of the TMIM setup include (1) high sensitivity (10 ~ 100 fm)[38] superior to that of commercial laser-based scanning vibrometers (~ 1 pm)[39], (2) outstanding spatial resolution (10 ~ 100 nm) well into the nanoscale, and (3) high operation frequency up to 10 GHz and beyond[26], making it an ideal technique to study integrated UHF/SHF phononic metamaterials.

We begin by characterizing the gap opening due to broken space-inversion symmetry. As illustrated in Fig. 2a, the PnC region is tilted from the normal of the plane wave launched by the IDT. Conservation of momentum parallel to the PnC boundary[14] requires that $k_\parallel = k_0 \sin \phi$, where $k_\parallel$ is the projection of $K$ (or $K'$) on the boundary, $k_0$ the wavevector for the pattern-free AlN membrane, and $\phi$ the tilt angle. As shown in Supplementary Information S2, this angular selection rule is satisfied at $\phi_1 = 23°$ (see Figs. 3 – 5) and $\phi_2 = 55°$ (see Fig. 2). In Figs. 2b and 2c, we plot the transmission spectra taken by a vector network analyzer on the gapless and gapped structures, respectively. Within the IDT passband (1.00 – 1.06 GHz), the transmission coefficient ($S_{21}$) of the gapless PnC is substantially higher (~ 30 dB) than that of the gapped PnC. The residual wave transmission in the gapped PnC is consistent with tunneling across the finite-sized (50 μm) crystal via an evanescent mode. The corresponding TMIM images at 1.045 GHz are shown in Figs. 2d and 2e. For the design with $\delta r = 0$, the gapless nature is evident from the appearance of transmitted plane waves on the opposite side of the PnC. In contrast, for the gapped structure with $\delta r = 100$ nm, the PnC region in the TMIM image (Fig. 2e) only displays the snowflake patterns due to topographic crosstalk (Supplementary Information S3), and the transmitted wave on the far side of the PnC is much weaker than for the gapless system (Fig. 2d). In addition, Fig. 2f shows the TMIM image inside the gapless PnC and the potential profile simulated by finite-element modeling. The resemblance between the two images confirms that the TMIM indeed measures the piezoelectric surface potential, which is proportional to the acoustic displacement field on the AlN membrane.

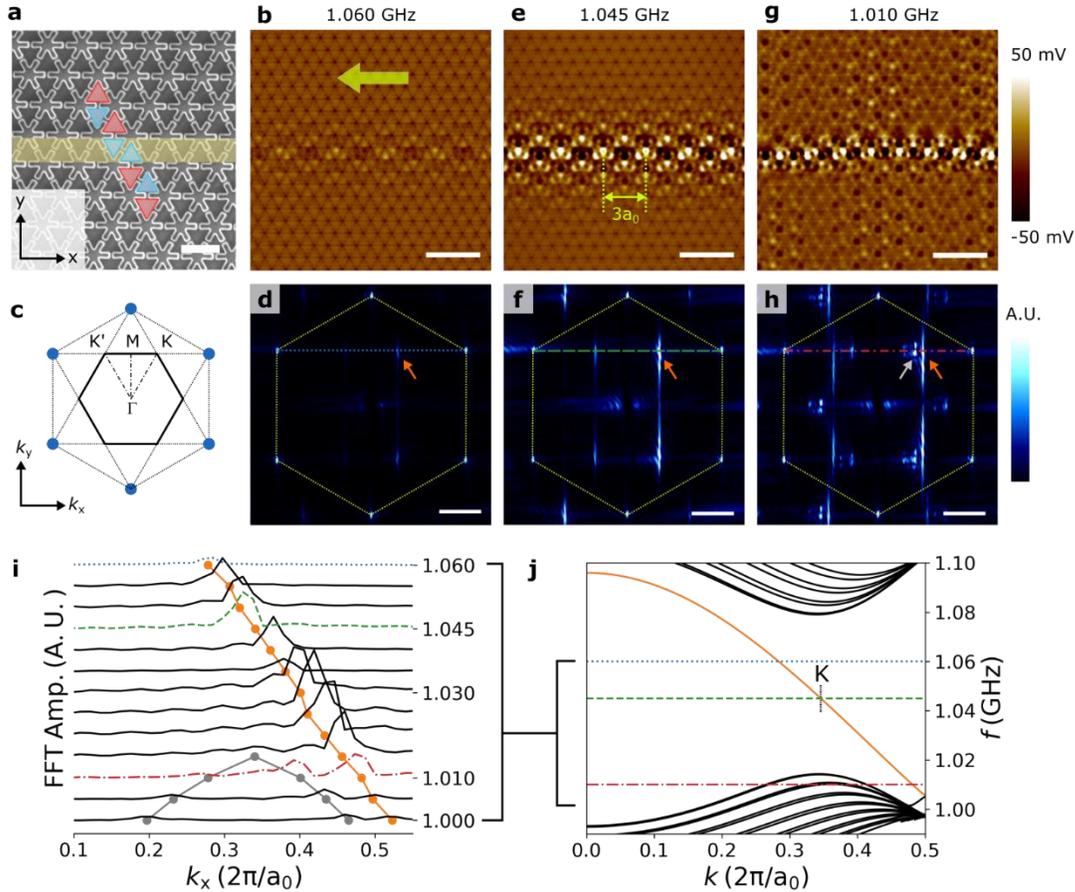

**Fig. 3 | Real-space imaging and momentum-space analysis of valley Hall edge states.**
**a**, SEM image near the interface (shaded in yellow) between two valley Hall insulators with $\delta r = +100$ nm (top) and $-100$ nm (bottom). The two sublattice sites are marked as blue and red triangles for clarity. **b**, TMIM image of the topological VHE state taken at $f = 1.060$ GHz. The arrow indicates that the elastic wave is launched from the right IDT and propagates in the $-x$-direction. **c**, Generic $k$-space map of the honeycomb lattice. The solid hexagon is the first Brillouin zone, with the high-symmetry points labeled in the map. The blue dots represent the reciprocal lattice sites. **d**, FFT amplitude map for the TMIM data at $f = 1.060$ GHz. The dashed hexagon is a guide for comparison with (**c**). The high-intensity line (denoted by the orange arrow) on the left side of the $K$ point is associated with the VHE state. **e**, TMIM image and **f**, FFT map at $f = 1.045$ GHz. The VHE wave pattern exhibits a period of $3a_0$. **g**, TMIM image and **h**, FFT map at $f = 1.010$ GHz. The bulk states are visible in both the real-space and $k$-space (denoted by the gray arrow) data. In (**d, f, h**), orange arrows point to the VHE states. **i**, Measured FFT amplitude along $KK'$ from 1 GHz to 1.06 GHz. The blue, green and red curves are extracted from the profiles indicate by the horizontal lines of corresponding colors in (**d, f, h**). The peak positions of the edge and bulk states are marked by orange and gray dots, respectively. **j**, Simulated bands of the PnC projected in the $k_x$ direction, showing the VHE band running across the bulk gap. The horizontal lines indicate the three representative frequencies of corresponding colors in **i**. The VHE edge state is colored in orange. Scale bars are 5 μm in (**a**), 20 μm in (**b, e, g**), and $2\pi \cdot 0.05$ μm$^{-1}$ in (**d, f, h**).

Having confirmed the symmetry-induced gap opening in our metamaterial design, we now move on to study the VHE edge states. Fig. 3a displays a scanning electron microscopy (SEM) image around the interface separating two domains with $\delta r = +100$ nm (top) and $-100$ nm (bottom). Since the difference in the topological charge between the two domains is quantized, i.e., $|\Delta C_K| = 1$, a chiral valley-Hall edge mode must exist at the boundary due to the bulk-edge correspondence[14]. Fig. 3b shows the real-space TMIM image at the upper bound of the IDT passband ($f = 1.060$ GHz), where a weak channel carrying elastic waves is observed (full data in Supplementary Information S4). The $k$-space information can be obtained by taking FFT of the signal $V_{\text{Ch1}} + i * V_{\text{Ch2}}$. Using the extended Brillouin zone of the honeycomb lattice (Fig. 3c) as a guide, we can discern a faint high-intensity line on the left side of the $K$ point (Fig. 3d). At $f = 1.045$ GHz (Fig. 3e), which coincides with the frequency at the $K$ point in reciprocal space, the VHE edge state is fully resolved as the frequency is well within the IDT band. Correspondingly, the FFT map in Fig. 3f exhibits very strong intensity at the $K$ point. According to the Bloch's theorem[40], the wavefunction of the edge state satisfies $\psi_{\vec{k}}(\vec{r}) = u_{\vec{k}}(\vec{r})e^{i\vec{k}\cdot\vec{r}}$, where $u_{\vec{k}}(\vec{r})$ is a function with the lattice periodicity. At the $K$ point, where $|k_x| = 2\pi/3a_0$, one should expect a period of $3a_0$ in the wavefunction, i.e., $\psi_K(x + 3a_0) = \psi_K(x)$, which is indeed observed in Fig. 3e. We emphasize that the weak intensity at the $K'$ valley in Fig. 3f is not due to intervalley scattering. Rather, it is associated with the inevitable reflection due to the impedance mismatch at the junction between the PnC region and the unpatterned region, which leads to wave propagation in the opposite direction (Supplementary Information S5). At $f = 1.010$ GHz, the frequency lies inside the bulk band such that both bulk and edge states are present (Fig. 3g). The $k$-space map in Fig. 3h also shows a high-intensity line on the right side of the $K$ point, which is associated with the VHE state, and some intensity around the $K$ point, which indicates the appearance of bulk states.

The high-resolution FFT maps allow us to quantitatively compare the experimental data and the calculated band structure. In the real space, the VHE edge state is a gapless mode localized at the interface and travels along opposite directions near $K$ and $K'$ valleys because of the opposite group velocity, manifesting the momentum-valley locking of the chiral edge states. The group velocity is characterized by the band dispersion as $d\omega/dk$. In Fig. 3i, we display the FFT line profiles along the $KK'$ direction and highlight the three representative frequencies shown in Figs. 3b-3h. The dispersion of both the bulk and topological edge states is clearly seen as we track the position of FFT peak intensity. This is in excellent agreement with the simulated curves in Fig. 3j, where the simulated bands projected in the $k_x$ direction are plotted. Here a single band appears around the $K$ valley and runs monotonically across the bulk bandgap of $1.015 - 1.080$ GHz.

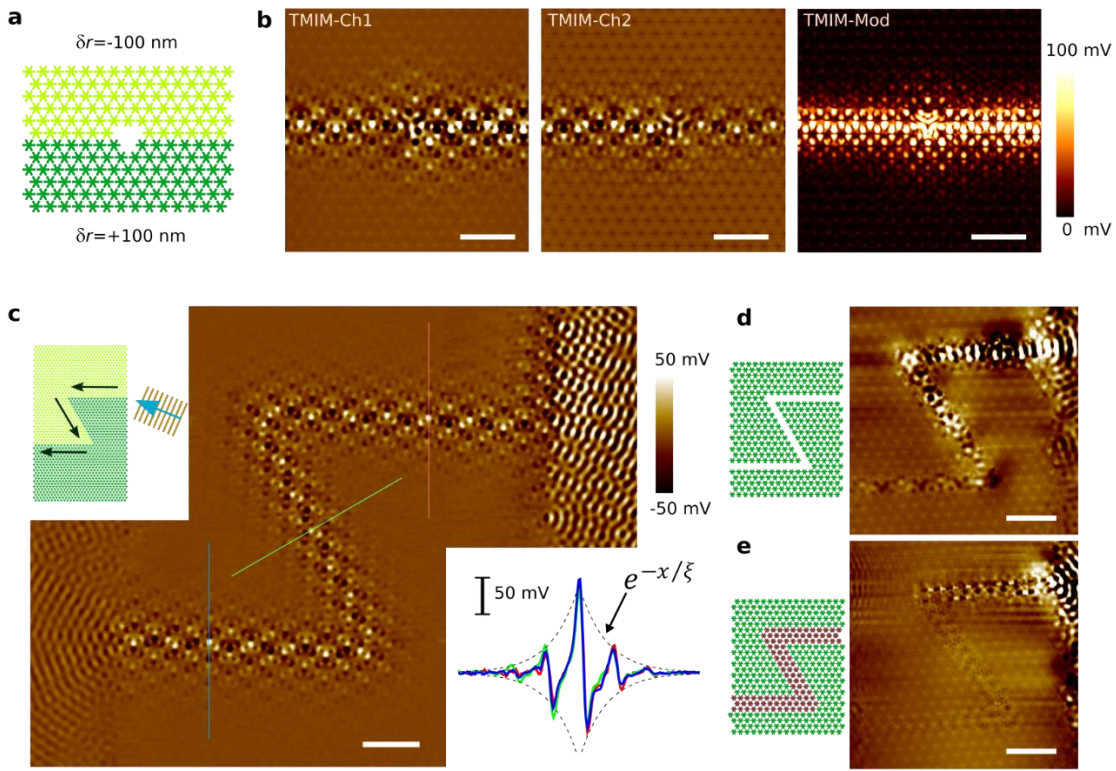

**Fig. 4 | Robustness of valley Hall edge transport. a**, Schematic of a local defect (three missing snowflakes) at the valley Hall interface. **b**, Left to right: TMIM-Ch1, -Ch2, and -modulus images of the sample in (**a**). The disorder alters the phase but not the amplitude of the elastic wave traveling along the topological boundary. **c**, TMIM image of a Z-shaped VHE edge channel (schematic in the top-left inset). The lower-right inset shows the line profiles at three representative locations denoted in the image. The dashed lines are exponential fits ($e^{-x/\xi}$) to the envelope function of the TMIM amplitude. **d**, TMIM images

of a Z-shaped pattern-free channel and **e**, a Z-shaped gapless channel embedded in gapped crystals. Schematics are shown to the left of the images. Substantial reflection and attenuation are evident for these topologically trivial interfaces. The false-color scale for all TMIM images is from −50 mV to 50 mV. All scale bars are 20 μm.

Using the same topological design, we can also demonstrate the robustness of VHE edge state transport against structural imperfections. Fig. 4a illustrates the schematic of a different PnC sample with one defect (three missing snowflakes) at the domain boundary[13]. As seen in Fig. 4b, while the signal strengths in both channels are altered by the scatterer, the TMIM modulus ($\sqrt{V_{\text{Ch}1}^2 + V_{\text{Ch}2}^2}$) remains unchanged after crossing the local disorder. In other words, the scattering center modifies the phase but not the amplitude of the elastic wave propagating along the topological interface. Similarly, the topological protection of valley transport is also validated by transmission with negligible loss through sharp corners of a zigzag interface[14], as illustrated in the inset of Fig. 4c. At the unit cell level, the relative position of the two valley Hall domains is invariant with respect to the interface. As a result, the forward-moving modes are always projected to the same valley and protected from backscattering by the band topology. As shown in the TMIM image (Fig. 4c), the incident elastic wave is guided into the domain wall and propagates freely through the two sharp turns. The absence of wave attenuation across the corners is evident from the TMIM signal profiles at three representative locations (lower right inset). The exponential decay ($e^{-x/\xi}$) of the wavefunction into gapped PnC regions indicates that the VHE state is indeed localized at the topological interface, with $\xi \sim 6$ μm comparable to the lattice constant. For comparison, we design and measure two topologically trivial Z-shaped channels – one pattern-free waveguide (Fig. 4d) and one gapless waveguide (Fig. 4e) – within gapped PnCs. In both cases, the first 120°-turn already introduces substantial reflection and strong disturbance to the incident wave. The attenuation after two bending corners is so significant that little energy is transmitted to the output port. The topological protection is thus crucial for guiding elastic waves with minimal loss in nanoelectromechanical systems.

Finally, we demonstrate the realization of a VHE-based topological beam splitter[33,41]. Fig. 5a shows the TMIM modulus image as an elastic wave enters a meta-structure with four alternating valley Hall domains (schematic in the upper right inset). The incident wave splits into two branches separated by an angle of 120°. On the other hand, the appearance of propagating wave along the straight route after the splitting would indicate that the wave is scattered to the other valley at the junction, which is prohibited by the VHE. In the experiment, wave transmission in the forward direction beyond the intersection is indeed negligible, with a suppression factor of over 40 dB. Interestingly, a careful examination of the crystal design (lower right inset of Fig. 5a) indicates that the upper branch is closer to the input line than the lower branch by $0.5a_0$. This subtle asymmetry is readily captured by the sensitive TMIM imaging. In Fig. 5b, we plot the square of TMIM modulus integrated over one wavelength ($\int_{a_0} (V_{Ch1}^2 + V_{Ch2}^2) dx$), which is proportional to the acoustic power, for all four segments. The lower branch receives less (~ 40%) acoustic energy than the upper branch (~ 60%) from the incident wave, which is consistent with the uneven beam splitting inherent in the metamaterial design.

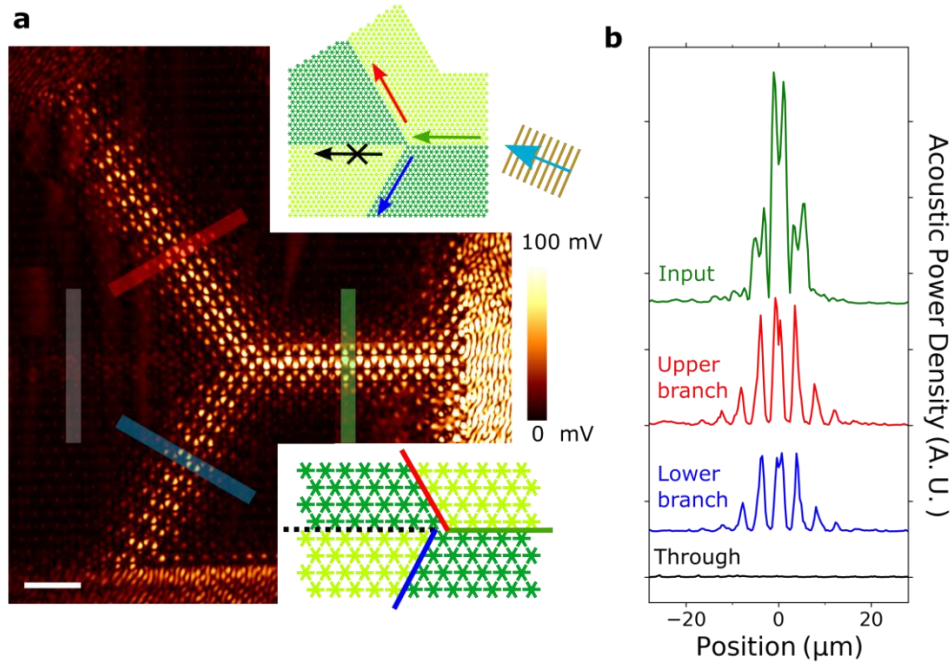

**Fig. 5 | Demonstration of topological beam splitting. a**, TMIM-modulus image of a topological beam splitter based on four alternating domains (schematic in the up-right inset). The lower-right inset shows a close-up view around the intersection. The scale bar is 20 μm. **b**, Square of TMIM-modulus signals integrated over one wavelength, which is proportional to the acoustic power, for the four paths labeled in (**a**).

The GHz topological VHE demonstrated in this work may be exploited for integrated phononic circuits in the UHF/SHF regime. The acoustic beam splitter in Fig. 5a, for instance, can be used as power dividers or combiners. The zigzag VHE edge channels in Fig. 4c allows robust 1D transport with small footprint, which is suitable for compact acoustic delay lines. Moreover, the valley-momentum locking property will enable us to implement valley filters. Here the gapless state ($\delta r = 0$) is used as the background PnC, where the valley degree of freedom is well defined. In this "acoustic graphene", both $K$ and $K'$ valley-polarized states can propagate in all directions allowed by the crystal symmetry. One can then construct a topological VHE region (similar to Fig. 3a) to select a specific valley state. After certain circuit operations, the valley information can be read out by passing it through a second VHE region that has either the same or opposite valley polarity. All these circuit elements are promising candidates for classical and quantum information applications.

In summary, we have designed and fabricated GHz nanoelectromechanical phononic crystals with topologically nontrivial structures. Using microwave impedance microscopy, we successfully visualize the elastic wave on patterned piezoelectric AlN membranes with unprecedented displacement sensitivity and spatial resolution. The topologically protected edge states between two gapped structures with opposite valley Chern numbers are demonstrated in both real-space imaging and momentum-space analysis. The valley Hall effect protected against backscattering is evident from the negligible loss through local disorders and sharp corners, as well as the power splitting into multiple edge channels. Our work provides an excellent framework to exploit integrated topological phononics for classical and quantum information processing in the microwave regime.

## Methods

**Device fabrication:** We fabricated the phononic structures on 800 nm thick $c$-axis polycrystalline AlN films grown by magnetron sputtering on Si wafers. The phononic crystal was formed by electron beam lithography (EBL) and plasma etching of AlN using 660-nm-thick $SiN_x$ hard mask. Interdigital transducers (IDTs) were used to emit and receive phonons. The IDT fingers have an aperture width of 80 μm. The IDTs were fabricated by EBL and lift-off of 45-nm-thick aluminum. Finally, the AlN film was released from the Si substrate by a $XeF_2$ etcher.

**Finite-element modeling:** The band structures and potential profiles are calculated using finite element methods in COMSOL Multiphysics 5.4. We compute the 2D and 1D band structures along with potential profiles in the corresponding geometries with periodic boundary conditions. The "piezoelectric effect Multiphysics" is applied which couples the solid mechanics module and electrostatics module. The material properties used are slightly modified from reported AlN membrane properties[42]. Specifically, $c_{11} = c_{22} = 375$ GPa , $c_{12} = 125$ GPa , $c_{13} = c_{23} = 120$ GPa , $c_{44} = c_{55} = 118$ GPa , $c_{66} = (c_{11} - c_{12})/2 = 125$ GPa , $e_{31} = e_{32} = -0.58$ C/m$^2$ , $e_{33} = 1.55$ C/m$^2$, $e_{15} = e_{24} = -0.48$ C/m$^2$, $\rho = 3180$ kg/m$^3$.

**Experimental setup:** The transmission-mode microwave impedance microscopy (TMIM) is implemented on an atomic-force microscopy platform (ParkAFM, XE-70). The shielded cantilever probe (Model 5-300N) is commercially available from PrimeNano Inc. At ~ 1 GHz, the cantilever probe can be viewed as a lumped element with an effective capacitance of ~ 1 pF. An impedance-match network is needed to route the tip to the 50-Ω transmission lines, resulting in a measurement bandwidth of ~ 70 MHz. In the experiment, this band is tuned to match the passband of the IDT. The tip/impedance-match receiver has an effective input impedance $|Z_{\text{in}}|$ ~ 1 kΩ at 1 GHz. Through a similar tip-sample coupling impedance $Z'_{\text{t-s}}$ ~ 100 kΩ, an input signal is picked up by the tip and then amplified and demodulated by the microwave electronics. Details of the TMIM experiment can be found in Ref. 25. All measurements are performed at room temperature.

**Data Availability.** All data supporting the findings of this study are available within the article and/or the SI Appendix. The raw data are available from the corresponding author upon reasonable request.

# Acknowledgements

This work was supported by the NSF through the Laboratory for Research on the Structure of Matter, an NSF Materials Research Science & Engineering Center (MRSEC; DMR-1720530). The TMIM work was supported by NSF Division of Materials Research Award DMR-2004536 and Welch Foundation Grant F-1814. The data analysis was partially supported by the NSF through the Center for Dynamics and Control of Materials, an NSF MRSEC under Cooperative Agreement DMR-1720595. This work was carried out in part at the Singh Center for Nanotechnology, which is supported by the NSF National Nanotechnology Coordinated Infrastructure Program under grant NNCI-2025608. The metamaterial design and simulation work was supported by the US Office of Naval Research (ONR) Multidisciplinary University Research Initiative (MURI) grant N00014-20-1-2325 on Robust Photonic Materials with High-Order Topological Protection and grant

## Author contributions

C.J. and K.L. conceived the project. Q.Z. fabricated the phononic devices and performed band-structure simulations. D.L. and L.Z. performed the TMIM imaging and data analysis. X.M. and S.I.M. contributed to the TMIM data analysis. L.H., Z.G., H.Y., and B.Z. contributed to the phononic crystal design. Q.Z., D.L., and K.L. drafted the manuscript with contributions from all authors. All authors have given approval to the final version of the manuscript.

## Competing interests

The authors declare no competing interests.

## Data availability

The data sets generated during the current study, and/or analysed during the current study, are available from the corresponding author upon reasonable request.